\documentclass[12pt]{article}
\usepackage{amssymb}
\usepackage{epsfig}

\catcode`\@=11
\def\numberbysection{\@addtoreset{equation}{section}
        \def\theequation{\thesection.\arabic{equation}}}

\def\beq{\begin{equation}}
\def\eeq{\end{equation}}
\numberbysection
\begin{document}
\begin{titlepage}
\begin{center}
\hfill DFF  1/11/03 \\
\vskip 1.in {\Large \bf Space-time noncommutativity and (1+1)
Higgs Model} \vskip 0.5in P. Valtancoli
\\[.2in]
{\em Dipartimento di Fisica, Polo Scientifico Universit\'a di Firenze \\
and INFN, Sezione di Firenze (Italy)\\
Via G. Sansone 1, 50019 Sesto Fiorentino, Italy}
\end{center}
\vskip .5in
\begin{abstract}
We compare the classical scattering of kinks in (1+1) Higgs model
with its analogous noncommutative counterpart. While at a
classical level we are able to solve the scattering at all orders
finding a smooth solution, at a noncommutative level we present
only perturbative results, suggesting the existence of a smooth
solution also in this case.
\end{abstract}
\medskip
\end{titlepage}
\pagenumbering{arabic}
\section{Introduction}

The concept of time in physics has been subject of endless
discussions. For Einstein time is equivalent to a spatial
coordinate, its role reduced to a mere parameter and the
distinction between past, present and future only an illusion of
the macroscopic level.

More recently time has emerged as a central theme in the problem
of irreversibility in both living and inanimate systems, as in
Prigogine's approach to nonequilibrium thermodynamics. For
Prigogine introducing in physics the concept of  "arrow of time"
\cite{1}-\cite{4}, irreversibility can have a constructive,
positive role creating order out of chaos. In this framework the
present-day laws of physics emerged from an initial chaotic
universe, where there was no law of physics at all, and the order
that we measure now is just an evolution under nonequilibrium
conditions due to the irreversibility of time.

However the physical community has always avoided discussing the
possibility of a microscopic irreversibility in quantum field
theory, conserving the spirit of Einstein. All the major physical
theories, i.e. quantum mechanics and general relativity, are
unitary by construction, in the sense that the quantum S-matrix
relating the "in" and "out" states must be unitary and conserve in
the scattering process the basis of the Hilbert space. Till now no
internal contradiction to this scheme has been found in ordinary
gauge theories and the experimental tests are coherent with it.

Problems arise when we try to unify quantum mechanics with general
relativity, i.e. when we perform ideal tests on the structure of
space-time. Hawking and Bekenstein in 1975 predicted that black
hole physics, taking into account quantum mechanics, is strictly
connected with thermodynamics, and that black holes can slowly
radiate. This prediction created an information loss problem, in
the sense that some information of the quantum system is lost
during the scattering process with a black hole.  Starting with a
quantum system in a pure state, the black hole is able to
transform it into a mixed state, described quantum-mechanically by
a density matrix rather than a wave function.  The transformation
of pure states into mixed states can be taken as a paradigm of
microscopic irreversibility, being a non-unitary process.

Quite recently an indetermination principle for space-time
coordinates has been suggested by a gedanken experiment
\cite{5}-\cite{7}, based on classical black holes and quantum
mechanics, pointing out to the study of quantum field theory on
noncommutative spacetimes \cite{8}-\cite{16}. An isomorphism
allows translating the noncommutativity of the coordinates to a
noncommutative star product of the fields on a commutative
space-time. This new type of nonlocality, especially in time, can
be very dangerous to the maintenance of the basic principles of
standard quantum field theory. In fact quantum field theories with
space-time noncommutativity have no straightforward Hamiltonian
quantization, which is usually the warranty for unitarity and
causality. Such theories are defined only through the Lagrangian
and via perturbation theory. To check unitarity in a Feynman
diagram it is enough to verify the cutting rules that relate the
amplitude of the Feynman diagram to its imaginary part. It has
been found that space-time noncommutativity indeed breaks the
standard cutting rules and therefore unitarity is lost
\cite{17}-\cite{18}. This fundamental result has been confirmed by
an analysis based on string theory propagating in an electric
background field. It is possible to restore unitarity in the
quantum field theory nonlocal in time by embedding it into string
theory, which is unitary, but the additional states, required for
recovering unitarity, cannot be decoupled in the field theory
limit. So it seems that there is no hope to maintain the structure
of standard quantum field theory in the case of space-time
noncommutativity \cite{17}-\cite{20}. We not able to judge other
approaches, based on axiomatic field theory, which promise that
restoring unitarity is possible modifying the definition of
quantum field theory on noncommutative space-times \cite{7}.

We instead want to push the idea that there is no mystery behind
these results, since the main physical motivation for introducing
noncommutative coordinates comes from black holes and the
Heisenberg indetermination principle, and black hole physics is
one of the best examples in physics where unitarity is lost and
the scattering process modifies the nature of quantum states, from
pure states to mixed ones. Therefore our interest in space-time
noncommutativity is motivated by interpreting it as a beautiful
model of microscopic irreversibility, which at least can be
controlled with a Lagrangian, while the black hole information
paradox is hard to study.

As a preliminary step, we have studied the classical scattering of
wave packets in a simple system with space-time noncommutativity
where perturbative resummations are possible, since we believe
that only nonperturbative results can be physically meaningful.
The system we have chosen is the Higgs model in ($1+1$)
dimensions, where classical solitons are the so-called kink
solutions. Since noncommutativity switches on only for an
interacting solution and the one-kink solution is trivial, we are
forced to study the scattering of two kinks, one left-mover
according to the equation $u = x-t = 0$, and the other right-mover
( $v = x + t = 0$ ).

We have simplified our ansatz of solution by taking two sharp
waves ( completely localized in space ) which can interact only
after $t>0$, when an extra interacting solution is needed to solve
the equations of motion. The simplicity of our ansatz allows us to
reduce the 2D equations of motion to a single nonlinear
differential equation depending on a single variable, which we are
able to solve exactly.

In the second part of the paper we attempt to generalize our
interacting classical solution to the case of noncommutative
kinks. Although we haven't been able to find the complete solution
we report our partial results postponing the exact solution to a
future research. In particular, the first-order correction in the
coupling constant is interesting because it modifies the profiles
of the two shock waves, which loose their sharpness and get a
dimension of order $\sqrt{\theta}$. Moreover the scattering
doesn't develop anymore at $t > 0$, but it is anticipated due to
the width of the wave packets.  Noncommutativity is able to modify
the asymptotic states permanently, also when the two wave packets
are far apart.  This result suggests that the rules of quantum
mechanics must be modified in presence of an infinite range force
( as noncommutativity appears to be ). However it would be more
interesting to interpret a complete solution, which we leave to a
future research.

\section{Higgs model for real scalar field}

Our aim was to find an exactly solvable scattering problem to
compare with the noncommutative case, possibly at a
nonperturbative level. Our choice has been to consider the
classical ($1+1$) Higgs model with a real scalar field defined by
the Lagrangian

\beq {\cal L} = \frac{1}{2} \partial_\mu \phi \partial_\mu \phi +
\frac{1}{2} m^2 \phi^2 - \frac{1}{4} \lambda \phi^4 \label{21}
\eeq

The Higgs mechanics is based on making perturbation theory around
the nontrivial minimum of the potential. To find it we need to
introduce the corresponding Hamiltonian :

\begin{eqnarray}
\pi & = & \frac{\partial {\cal L}}{\partial ( \partial_0 \phi )} =
\partial_0 \phi \nonumber \\
{\cal H} & = & \frac{1}{2} ( \pi^2 + | \partial_i \phi |^2 ) +
V(\phi) \nonumber \\
V(\phi) & = & \frac{1}{4} \lambda \phi^4 - \frac{1}{2} m^2 \phi^2
\label{22}
\end{eqnarray}

The minimum of the potential (\ref{22}) is reached when

\beq \frac{\partial V(\phi)}{\partial \phi} = 0 \rightarrow \phi^2
= \frac{m^2}{\lambda} \label{23} \eeq

Let us choose for simplicity the coupling constants such that
$m=\sqrt{\lambda}$ and $\phi = \pm 1$ are the two nontrivial
minima.

The corresponding equations of motion

\beq \square \phi = \lambda \phi ( 1 - \phi^2 ) \label{24} \eeq

can be simplified by defining the light-cone variables :

\begin{eqnarray}
& \ & u = \frac{x-t}{2} \ \ \ \ \ v = \frac{x+t}{2} \nonumber \\
& \ &  \square = - \partial_u \partial_v \nonumber \\
& \ &  \partial_u \partial_v \phi = - \lambda \phi ( 1 - \phi^2 )
\label{25}
\end{eqnarray}

Analyzing such equation one notice that $ \phi = 0, \pm 1$ are
possible solutions. Let us generalize them by introducing as an
ansatz a sum of step functions which interpolate between the three
values $0, \pm 1$:

\beq \phi^{(0)} = - 1 + p ( \Theta (u) - \Theta (v) )  \ \ \ \  p
= 1,2 \label{26} \eeq

Due to the dependence from single light-cone variables,
$\phi^{(0)}$ is certainly a zero of the first member of the
equation (\ref{25}), however the second member is null only if $ t
< 0$. For $t>0$ the ansatz $\phi^{(0)}$ is no longer valid and
must be generalized to

\beq \phi = - 1 + p ( \Theta (u) - \Theta (v) ) + f(uv) \Theta(v)
\Theta(-u) \label{27} \eeq

where the unknown function f is dependent on the single variable
uv \footnote[1]{To avoid confusion we define the step functions as
$\Theta (u) = 1 \ {\rm if} \ u > 0, \ {\rm or} \ 0 \ {\rm
otherwise}, \Theta (v) = 1 \ {\rm if} \ v > 0, \ {\rm or} \ 0 \
{\rm otherwise}$, and $\Theta(-u) \equiv 1 - \Theta (u)$}. This
interacting solution must evolve such that asymptotically, at
infinite time and finite space, i.e. for $uv \rightarrow -
\infty$, the function reaches a constant value. The particular
value  $f(-\infty)$ is determined by solving the equations of
motion (\ref{25}).

The solution with $p=1$ has as "in state" $\phi_{in} = 0$ ( for $t
\rightarrow - \infty$ ) , while the "out state" ( for $ t
\rightarrow +\infty$ ) has to be determined by solving the
equations of motion (\ref{25}), and it is parameterized as
$f_{p=1} ( -\infty) - 2 $.

The solution with $p=2$ has as "in state" $\phi_{in} = 1$ ( for $t
\rightarrow - \infty$ ) and the "out state" $f_{p=2} ( -\infty)
-3$.

In synthesis

\begin{eqnarray}
p & = & 1 \ \ \phi_{in} = 0 \ \ \ \rightarrow \ \phi_{out} =
f_{p=1} (-\infty) - 2 \nonumber \\
p & = & 2 \ \ \phi_{in} = 1 \ \ \ \rightarrow \ \phi_{out} =
f_{p=2} (-\infty) - 3 \label{28}
\end{eqnarray}

By introducing the complete ansatz (\ref{27}) into the equations
of motion (\ref{25}) one is able to show that the ansatz closes if
all the following conditions are met:

\begin{eqnarray}
& \ & f(0) = 0 \ \  \ f' (uv) \ \ {\rm regular \ around } \ uv \sim 0 \nonumber \\
& \ & (uv) f''(uv) + f'(uv) = \lambda ( f(uv)-p ) ( f(uv)-p-1 ) (
f(uv)-p-2 )
\nonumber \\
& \ & p ( p-1 ) ( p-2 ) = 0 \label{29}
\end{eqnarray}

The ansatz closes for the values $p=0,1,2$; since the value $p=0$
is trivial, in the following we will discuss only the solution to
the differential equation (\ref{29}) for the values $p = 1,2$.

By introducing the variable

\beq x = -  6 \lambda uv \label{210} \eeq

we are led to discuss the solution to the following equation:

\beq x f''(x) + f'(x) + ( f(x)-p )( f(x)-p-1 )( f(x)-p-2 )/6 = 0
\label{211} \eeq

in the range $0 < x < +\infty$.

We will show that the nonlinear equation (\ref{211}) is consistent
with the following boundary values:

\begin{eqnarray}
& \ & f_{p=1} ( x = + \infty ) = 1 \nonumber \\
& \ & f_{p=2} ( x = + \infty ) = 4 \label{212}
\end{eqnarray}

leading to classify the possible scenarios:

\begin{eqnarray}
& \ & p = 1 \ \ \ \ \phi_{in} = 0 \ \ \ \rightarrow \ \ \
\phi_{out} = -1 \nonumber \\
& \ & p = 2 \ \ \ \ \phi_{in} = 1 \ \ \ \rightarrow \ \ \
\phi_{out} = 1 \label{213}
\end{eqnarray}

In the first case, the scattering of kinks allows to describe the
decay from the unstable state $\phi_{in} = 0$ to the stable
minimum of the potential $\phi_{out} = -1$; in the second case,
the scattering of kinks doesn't alter the stability of the minimum
$\phi_{in} = \phi_{out} = 1$.

Let us start to solve (\ref{211}) with $p=1$. In this case we
define

\begin{eqnarray}
& \ &  f(x) = 1 - g(x) \ \ \ \ g(0) = 1 \nonumber \\
& \ & x g''(x) + g'(x) +  g(x) ( g(x) + 1 ) ( g(x) + 2 ) / 6 = 0
\label{214}
\end{eqnarray}

Being a second order differential equation, it seems that the only
boundary value $g(0) = 1$ is not enough to determine completely
the solution, but it turns out that another physical requirement
is necessary to obtain a smooth solution, i.e. the absence of
logarithmic terms around $x=0$. The request is sufficient to
determine the asymptotic vale $g(+\infty) = 0$, which can be
achieved or by a direct numerical computation with Mathematica, or
with a careful inspection of the differential equation. We have
done both checks and they completely agree.

Firstly, let us suppose that asymptotically $ g(x) \rightarrow 0$
for $ x \rightarrow +\infty $, then the nonlinearity can be
avoided and the nonlinear problem (\ref{211}) can be linearized to

\begin{eqnarray}
& \ & x g''(x) + g'(x) +  g(x) / 3  = 0 \nonumber \\
& \ & g(0) = 1 \ \ \ \ g(x) {\rm \ \ regular \  around \ } x = 0
\label{215}
\end{eqnarray}

which can be solved by the Bessel function:

\beq g(x) = J_0 ( 2\sqrt{\frac{x}{3}} ) \label{216} \eeq

Since it is well known the asymptotic behaviour of the Bessel
function

\beq \lim_{x \rightarrow +\infty} J_0 (x) = \sqrt{ \frac{2}{\pi x}
} cos ( x - \frac{\pi}{4} ) \label{217} \eeq

we conclude that the asymptotic value $ g(+\infty) = 0 $ is
consistent with the hypothesis, and we have learned that the
solution of (\ref{211}) has asymptotic damped oscillations
behaving as $ x^{-\frac{1}{4}}$.

The difference between linear and nonlinear solutions is
concentrated around $x = 0$; however supposing that the value of
$g(x)$ is always greater than $-1$, at the minima where $g'(x)
=0$, the value of the second derivative $g''(x)$ is always
opposite to the value of the function $g(x)$, confirming the
oscillating character of the solution around the value $g = 0$,
also at the nonlinear level. At this point, we are ready to
compare these considerations with a direct numerical computation
with Mathematica.

We chose the following method. Firstly we  developed in power
series the solution around zero and put the recurrence relations
for the coefficients of the series into Mathematica. The power
series truncated at, let's say, 100 steps has a certain
convergence radius, inside it we can trust the approximated values
of the function $g(x)$. Then we used these approximated values to
build the recurrence relations around another fixed point inside
the convergence radius of the first power series, and we iterated
the procedure. In this way we have found the locus of the first 6
minima ( these results should be taken as indicative values due to
the imprecision of the extrapolated values )

\begin{eqnarray}
 & \ & x = 11  \ \ \ \ \  g(x) = -0.62 \nonumber \\
 & \ & x = 90 \ \ \ \ \ g(x) = - 0.49 \nonumber \\
 & \ & x = 244 \ \ \ \ \ g(x) = - 0.36 \nonumber \\
 & \ & x = 462 \ \ \ \ \ g(x) = - 0.32 \nonumber \\
 & \ & x = 745 \ \ \ \ \ g(x) = - 0.28 \nonumber \\
 & \ & x = 1089
\ \ \ \ g(x) = - 0.25 \label{218}
\end{eqnarray}

and of the first 5 local maxima:

\begin{eqnarray}
& \ & x = 41 \ \ \ \ \  g(x) = 0.37 \nonumber \\
& \ & x = 161 \ \ \ \ \ g(x) =  0.29 \nonumber \\
& \ & x = 346 \ \ \ \ \ g(x) =  0.25 \nonumber \\
& \ & x = 598 \ \ \ \ \ g(x) =  0.23 \nonumber \\
& \ & x = 911 \ \ \ \ \ g(x) =  0.21 \label{219}
\end{eqnarray}

In Fig.1 we have plotted with Mathematica two curves representing
the approximated value of the function $g(x)$ calculated from the
power series truncated at an even and odd number respectively (
for example 100 and 99 steps ):

\begin{figure}[htbp]
  \begin{center}
    \includegraphics[width = 6.5cm]{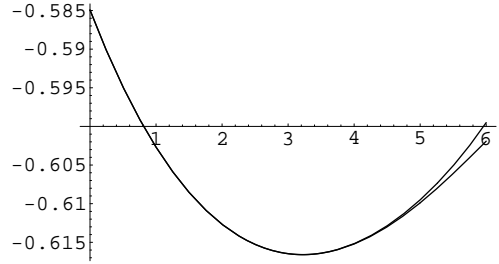}
    \caption{The first minimum for $p=1$} \label{grafico1}
  \end{center}
\end{figure}

By analyzing these results we find agreement with all the
preliminary discussion, since the minimum value of $g(x)$
($-0.62$) is greater than $-1$, from which damped oscillations
follow until reaching the linear behaviour (\ref{217}). Thus by
combining numerical and analytic methods we have full control of
the nonlinear equation (\ref{214}).

Let us now discuss the solution to the equations of motion for
$p=2$, in which case the final state of the Higgs field
$\phi_{out} = f_{p=2} (-\infty) - 3$. By defining $f(x) = 2 (1 -
g(x)) $ and rescaling $ x \rightarrow 2x $, we obtain the
following differential equation for $g(x)$:

\beq x g''(x) + g'(x) + g(x) ( 1 + g(x) ) ( 1 + 2 g(x) )/6 = 0
\label{220} \eeq

At a first sight this equation looks very similar to the one
discussed before, but in reality its solution is quite different.
Firstly we notice that it is not clear what is the final point of
oscillation. There are at least two possible choices:

i) $g(x)$ oscillates around the value $g=0$, then to be
self-consistent, at the stationary points, the value of the second
derivative must be opposite to the value of the function and this
happens if the function is confined over the minimum value  -1/2 .

ii) $g(x)$ oscillates around the value $ g = - 1$; this is
possible if the local maxima and minima are confined under the
maximum value -1/2.

Only with a numerical computation we have been able to discern the
right value. By using the same method illustrated before we have
found the locus of the first 4 minima

\begin{eqnarray}
 & \ & x = 41  \ \ \ \ \  g(x) = -1.14 \nonumber \\
 & \ & x = 212 \ \ \ \ \ g(x) = - 1.10 \nonumber \\
 & \ & x = 506 \ \ \ \ \ g(x) = - 1.08 \nonumber \\
 & \ & x = 922 \ \ \ \ \ g(x) = - 1.07
\label{221}
\end{eqnarray}

and of the first 4 local maxima:

\begin{eqnarray}
& \ & x = 108 \ \ \ \ \  g(x) = - 0.85 \nonumber \\
& \ & x = 342 \ \ \ \ \ g(x) =  - 0.89 \nonumber \\
& \ & x = 697 \ \ \ \ \ g(x) =  - 0.91 \nonumber \\
& \ & x = 1173 \ \ \ \ \ g(x) =  - 0.92 \label{222}
\end{eqnarray}

\begin{figure}[htbp]
  \begin{center}
    \includegraphics[width = 6.5cm]{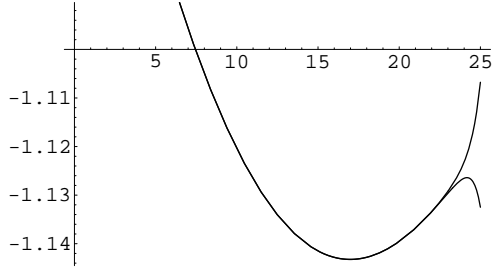}
    \caption{The first minimum for $p=2$} \label{grafico2}
  \end{center}
\end{figure}

We conclude that the possibility ii) is realized, and therefore
$g(x)$ has as asymptotic value -1, with oscillations that are
damped by a factor $x^{-1/4}$, the typical factor of the Bessel
function (\ref{217}). This completes the demonstration of the
boundary values depicted in (\ref{212}).

\section{Noncommutative case}

We are going to deform the Higgs model with a noncommutative
relation between the coordinates, for example

\beq [u,v] = \ i  \theta \  \ \ \leftrightarrow \ \ \ [x,t] = \ 2
i \theta \label{31} \eeq

This can be accomplished by deforming the ordinary product of
fields into an associative star product as follows:

\beq \phi_1 (u,v) * \phi_2 (u,v) = \lim_{u_1 \rightarrow u_2}
\lim_{v_1 \rightarrow v_2} e^{\ i \frac{\theta}{2} (
\partial_{u_1}
\partial_{v_2} - \partial_{u_2} \partial_{v_1} )} \ \phi_1 ( u_1,
v_1) \ \phi_2 (u_2, v_2) \label{32} \eeq

The lagrangian of the Higgs field with a noncommutative star
product is defined as:

\beq {\cal L} = \int d^2 x \left( \frac{1}{2} \partial_\mu \phi
\partial_\mu \phi + \frac{1}{2} m^2 \phi^2 - \frac{\lambda}{4}
\phi * \phi * \phi * \phi \right) \label{33} \eeq

and the equations of motion ( with $m^2/\lambda =1$ ) are nonlocal
in time:

\beq \partial_u \partial_v \phi = \lambda \phi * ( \phi * \phi -1
) \label{34} \eeq

It is not clear at this point how to complete the ansatz
(\ref{26})

\beq \phi^{(0)} = -1 + p ( \Theta (u) - \Theta (v) ) \label{35}
\eeq

in order to reduce the equations of motion (\ref{34}) into a
self-consistent set of equations. We prefer to setup a
perturbative method. In this framework the natural perturbative
parameter is the coupling constant $\lambda$, and we develop the
solution $\phi$ as a sum

\beq \phi = \phi^{(0)} + \lambda \phi^{(1)} + O(\lambda^2)
\label{36} \eeq

Starting from the ansatz $\phi^{(0)}$ (\ref{35}), we calculate the
source (\ref{34}) at the first perturbative order in $\lambda$. In
the case of the ordinary Higgs model we would obtain

\begin{eqnarray} \partial_u \partial_v \phi^{(1)} & = & \lambda \phi^{(0)} ( \phi^{(0) 2} -1 )
\nonumber \\ & = & \lambda p (p-1)(p-2) ( \Theta (u) - \Theta (v)
) - 6 \lambda p^2 \Theta (-u) \Theta (v) \label{37}
\end{eqnarray}

For the special cases $p=1,2$ the first term in the second member
cancels out and the main contribution comes from the term $\Theta
(-u) \Theta (v)$ which is different from zero only for $t > 0$,
and in the interval $ -t < x < t$.

The interacting field $\phi^{(1)}$ is then proportional to $ x^2 -
t^2 $

\beq \phi^{(1)} = - 6 \lambda p^2 uv \Theta (-u) \Theta (v)
\label{38} \eeq

In the noncommutative case we limit ourself to a calculation of
$\phi^{(1)}$ always starting from the ansatz $\phi^{(0)}$ but
replacing the ordinary product with the star product

\beq \partial_u \partial_v \phi^{(1)}_{NC} = \lambda \phi^{(0)} *
( \phi^{(0)} * \phi^{(0)} - 1 ) \label{39} \eeq

However we encounter the first difficulty, i.e. the star product
involving step functions seems to be ill-defined, being a sum of
infinite distributions. We will do the following trick, i.e.
solving the star product using the Fourier transform.

As an exercise, let us calculate the ordinary product $\phi^{(0)
2}$ using the Fourier transform of $\phi^{(0)}$. It is more
convenient to perform the Fourier transform of $\phi^{(0)} (u,v)$
with respect to the single variable u:

\begin{eqnarray} \tilde{\phi} ( k,v) & = & \frac{1}{\sqrt{2\pi}} \int^{+\infty}_{-\infty}
du e^{ - i k u} \phi^{(0)} ( u, v) = \nonumber \\
& = & \sqrt{ 2\pi} \left[ - ( 1 + p \Theta (v) ) \delta (k) +
\lim_{\epsilon \rightarrow 0} \frac{ p }{ 2 \pi i ( k - i \epsilon
)} \right] \label{310}
\end{eqnarray}

The ordinary product $\phi^{(0) 2}$ is mapped , under the Fourier
transform , to a convolution product:

\begin{eqnarray} \phi^{(0) 2} ( u,v) & \rightarrow & \frac{1}{\sqrt{2\pi}}
\int^{+\infty}_{-\infty} d q \tilde{\phi} ( q, v) \tilde{\phi} ( k - q , v ) = \nonumber \\
& = & \sqrt{2\pi} \left[ ( 1 + p ( p+2 ) \Theta (v) ) \delta (k) +
\lim_{\epsilon \rightarrow 0}  \frac{ p ( p-2 ) - 2 p^2 \Theta (v)
}{ 2 \pi i ( k - i \epsilon ) }\right] \label{311}
\end{eqnarray}

which is exactly the Fourier transform of

\beq \phi^{(0) 2} (u,v) = 1 + p ( p + 2 ) \Theta (v) + ( p ( p - 2
) - 2 p^2 \Theta (v) ) \Theta (u) \label{312} \eeq

The tool that we need during the calculation of the noncommutative
case is how the star product of two functions of u and v is
translated into a convolution product of their Fourier transforms.
It is not difficult to show that $\phi_1 * \phi_2 $ , defined the
Fourier transforms

\begin{eqnarray}
\phi_1 ( u_1, v_1 ) & = & \frac{1}{\sqrt{ 2 \pi}}
\int^{+\infty}_{-\infty} d q_1 \ e^{i q_1 u_1}
\ \tilde{\phi}_1 ( q_1, v_1 ) \nonumber \\
\phi_2 ( u_2, v_2 ) & = & \frac{1}{\sqrt{ 2 \pi}}
\int^{+\infty}_{-\infty} d q_2 \ e^{i q_2 u_2} \ \tilde{\phi}_2 (
q_2, v_2 ) \label{313}
\end{eqnarray}

is mapped to the following NC convolution product

\beq \widetilde{ ( \phi * \phi )} ( k, v) = \frac{1}{\sqrt{ 2
\pi}} \int^{+\infty}_{-\infty} dq \ \tilde{\phi}_1 \left( q, v +
\frac{\theta}{2} ( k-q ) \right) \ \tilde{\phi}_2 \left( k - q, v
- \frac{\theta}{2} q  \right) \label{314} \eeq

We notice from this definition that if the two functions are equal
( $\phi_1 = \phi_2$ ), then, due to the symmetry of the
convolution product $ q \rightarrow k-q $, the dependence on the
$\theta$ variable becomes even, as it happens for the star product
of the same function.

Let us calculate the star product $\phi^{(0)} * \phi^{(0)}$ using
the rule (\ref{314})

\begin{eqnarray} \widetilde{ ( \phi^{(0)} * \phi^{(0)} ) } ( k,v ) & = & \sqrt{ 2 \pi }
[ \ ( 1 + p ( p + 2 ) \Theta (v) \delta (k) +  \nonumber \\
& + &  \lim_{\epsilon \rightarrow 0} \frac{ p ( p-2 ) - p^2 (
\Theta ( v + \frac{\theta}{2} k ) + \Theta ( v - \frac{\theta}{2}
k ) ) }{ 2 \pi i ( k - i \epsilon ) } \ ] \label{315}
\end{eqnarray}

Therefore the pure noncommutative contribution is

\beq \widetilde{ ( \phi^{(0)} * \phi^{(0)} ) }_{NC} ( k,v )
 = - \frac{p^2}{\sqrt{ 2 \pi} i } \lim_{\epsilon \rightarrow 0}
 \left( \frac{\Theta ( v + \frac{\theta}{2} k ) +
\Theta ( v - \frac{\theta}{2} k ) - 2 \Theta (v)}{k-i \epsilon}
\right) \label{316} \eeq

In the following we will assume for simplicity that $uv >0$ and
$\theta > 0$, otherwise some signs function should to be added,
with the result of making the notations heavier.

Let us perform the anti-Fourier transform of (\ref{315}) to have a
better idea of the noncommutative source:

\begin{eqnarray} ( \phi^{(0)} * \phi^{(0)} )_{NC} & = & \int^{\theta}_0 d\theta
\frac{\partial}{\partial \theta} {( \phi^{(0)} * \phi^{(0)}
)}_{NC} = \frac{1}{\sqrt{2\pi}} \int^{\theta}_0 d \theta \
\int^{+\infty}_{-\infty} dk \ e^{i k u} \
\frac{\partial}{\partial \theta} \widetilde{( \phi^{(0)} * \phi^{(0)} )}_{NC} = \nonumber \\
& = &  - \frac{p^2}{\pi} \int^{\frac{2uv}{\theta}}_{+\infty} dx
\frac{ sin x}{x} = - \frac{p^2}{\pi} \left[ {\cal S}i \left(
\frac{ 2uv }{\theta} \right) - {\cal S}i ( + \infty ) \right]
\label{317}
\end{eqnarray}

This function is a special function, known as the Sine Integral,
defined as

\beq {\cal S}i (z) = \int^z_0 dx \ \frac{ sin x }{x} \ \ \ \ \ \ \
\ \ \ {\cal S}i ( + \infty )  = \frac{\pi}{2} \label{318} \eeq

Therefore we conclude that the noncommutative part of the source
has support only in the small region $ 0 < uv \lesssim \theta $
(otherwise the contribution is negligible) around the sharp wave
packets, and its role is to give a size of order $\sqrt{\theta}$
to the noncommutative kinks.

The next step is to perform the complete calculation of the source
(\ref{39}). We will compare again the commutative case with the
noncommutative one, to be able to extract the pure noncommutative
part. In the commutative case we need to compute in Fourier
transform the product $ \phi^{(0)} \  ( \phi^{(0} \phi^{(0)} - 1 )
$ where

\begin{eqnarray}
& \ & \tilde{\phi}^{(0)} = \sqrt{2\pi} \ \left[ \ - ( 1 + p \Theta
(v) ) \delta (k) +
\lim_{\epsilon \rightarrow 0} \frac{p}{2\pi i ( k- i \epsilon)} \ \right] \nonumber \\
& \ & \widetilde{( \phi^{(0)} \phi^{(0)} - 1 )} = \sqrt{2\pi} \
\left[ \ p ( p + 2 ) \Theta (v) \delta (k) + \lim_{\epsilon
\rightarrow 0} \frac{ p ( p - 2 ) - 2 p^2 \Theta (v) }{ 2 \pi i  (
k - i \epsilon ) } \ \right] \label{319}
\end{eqnarray}

We obtain

\begin{eqnarray}
& \ & \widetilde{ \phi^{(0)} \ ( \phi^{(0)} \phi^{(0)} - 1 )} ( k,
v ) = \sqrt{2\pi} \
[ \ - p ( p + 1 ) ( p + 2 ) \Theta (v) \delta (k) + \nonumber \\
& \ & + \lim_{\epsilon \rightarrow 0} \frac{ p ( p-1 ) ( p-2 ) + 6
p^2 \Theta (v)}{ 2\pi i ( k - i \epsilon )} \ ] \label{320}
\end{eqnarray}

It is easy to verify the correctness of this result, remembering
that

\beq \phi^{(0)} ( \phi^{(0) 2} - 1 ) = - p ( p + 1 )( p + 2 )
\Theta (v) + p ( p-1 ) ( p-2 ) \Theta (u) + 6 p^2 \Theta (u)
\Theta (v) \label{321} \eeq

Let us compute the noncommutative case:

\beq \phi^{(0)} * ( \phi^{(0)} * \phi^{(0)} - 1 ) \label{322} \eeq

where

\begin{eqnarray} \widetilde{ ( \phi^{(0)} * \phi^{(0)} - 1 ) }  & = & \sqrt{2\pi} \ [ \ p ( p + 2 )
\Theta (v) \delta (k) + \frac{1}{2\pi i } \lim_{\epsilon \rightarrow 0}
\frac{ p ( p-2 )}{ k - i \epsilon } \nonumber \\
& \ & - \lim_{\epsilon \rightarrow 0} \frac{ p^2 ( \Theta ( v +
\frac{\Theta}{2} k ) + \Theta ( v - \frac{\Theta}{2} k ) )} { 2
\pi i ( k - i \epsilon ) } \  ] \label{323}
\end{eqnarray}

After simple but tedious calculations we arrive at the following
result

\begin{eqnarray}
\widetilde{ \phi^{(0)} * ( \phi^{(0)} * \phi^{(0)} - 1 ) } ( k, v)
& = & \sqrt{ 2\pi} \ \left[ \ - p ( p+1 ) ( p+2 ) \Theta (v)
\delta (k) +
\lim_{\epsilon \rightarrow 0}
\frac{ p ( p-1 ) ( p-2 )}{ 2\pi i ( k- i \epsilon )} \right. \nonumber \\
& + &  \lim_{\epsilon \rightarrow 0} \frac{ p^3 ( \Theta ( v +
\frac{\theta}{2} k )  \Theta ( v - \frac{\theta}{2} k )
- \Theta (v) ) }{ 2 \pi i ( k - i \epsilon ) } \nonumber \\
& + & \lim_{\epsilon \rightarrow 0} \frac{ 3 p^2 ( \Theta ( v +
\frac{\theta}{2} k ) + \Theta ( v - \frac{\theta}{2} k ) )}{
 2 \pi i ( k - i \epsilon ) } \nonumber \\
& - & \left. \frac{p^3}{2 ( 2 \pi i )^2} \lim_{\epsilon
\rightarrow 0} \int^{+\infty}_{-\infty} dq \frac{ \Theta ( v +
\theta q ) + \Theta ( v - \theta q ) - 2 \Theta (v) }{ ( q +
\frac{k}{2} - i \epsilon )
 ( \frac{k}{2} - q - i \epsilon )} \right]
\label{324}
\end{eqnarray}

Let us extract the pure noncommutative part which is made of three
parts:

\beq \widetilde{ \phi^{(0)} * ( \phi^{(0)} * \phi^{(0)} - 1 )
}_{NC} ( k, v, \theta) = I_1 ( k, v, \theta) + I_2 ( k, v, \theta)
+ I_3 ( k, v, \theta) \label{325} \eeq

where
\begin{eqnarray}
& \ & I_1 ( k, v, \theta) = \lim_{\epsilon \rightarrow 0} \frac{ 3
p^2 ( \Theta ( v + \frac{\theta}{2} k ) + \Theta ( v -
\frac{\theta}{2} k ) - 2 \Theta (v) )}{
\sqrt{2 \pi} i ( k - i \epsilon ) } \nonumber \\
& \ & I_2 ( k, v, \theta) = \lim_{\epsilon \rightarrow 0} \frac{
p^3 ( \Theta ( v + \frac{\theta}{2} k )  \Theta ( v -
\frac{\theta}{2} k )
- \Theta (v) ) }{ \sqrt{2 \pi} i ( k - i \epsilon ) } \nonumber \\
& \ & I_3 ( k, v, \theta) = - \frac{p^3 \sqrt{2\pi} }{2 ( 2 \pi i
)^2} \lim_{\epsilon \rightarrow 0}  \int^{+\infty}_{-\infty} dq
\frac{ \Theta ( v + \theta q ) + \Theta ( v - \theta q ) - 2
\Theta (v) }{ ( q + \frac{k}{2} - i \epsilon )
 ( \frac{k}{2} - q - i \epsilon )}
\label{326}
\end{eqnarray}

Let us compute the anti-Fourier transforms of $I_i$:

\beq I_i ( u, v ) = \frac{1}{\sqrt{2\pi}} \int^{\theta}_0 d \theta
\  \int^{\infty}_{-\infty} dk \ e^{i k u} \
\frac{\partial}{\partial \theta} I_i ( k, v, \theta ) \label{327}
\eeq

With the trick of integrating and deriving with respect to
$\theta$, all the integrals can be done, obtaining the following
results:

\beq  { \phi^{(0)} * ( \phi^{(0)} * \phi^{(0)} - 1 ) }_{NC} ( u,
v, \theta) = \frac{ 3 p^2 + p^3 ( \Theta (v) - \Theta (u) )}{\pi}
\left[ {\cal S}i \left( \frac{2uv}{\theta} \right) - {\cal S}i ( +
\infty ) \right] \label{328} \eeq

Let's start integrating this source by taking the position

\beq \phi^{(1)}_i ( u, v) = \frac{ 3 p^2 + p^3 ( \Theta (v) -
\Theta (u) )}{2\pi} \ \int^{\theta}_0 d\theta \ \phi^{(1)} \left(
\frac{2uv}{\theta} \right) \label{329} \eeq

where

\beq \partial_u \partial_v \phi^{(1)} \left( \frac{2uv}{\theta}
\right) = - \frac{2 \lambda}{\theta} sin \left( \frac{2uv}{\theta}
\right) \label{330} \eeq

Let us define $x = 2uv / \theta $, then equation (\ref{330}) is
equivalent to

\beq \partial_x  (  x \partial_x \phi^{(1)} (x)  ) = - \lambda sin
x \ \ \ \ \ \rightarrow \ \ \ \ \phi^{(1)} (x) =  \lambda {\cal
C}i \left( \frac{2uv}{\theta} \right) \label{331} \eeq

where we have introduced the Cosine Integral function defined as:

\beq {\cal C}i (z) = \int^z_{+\infty} dx \ \frac{ cos x }{x}
\label{332} \eeq

However the steps functions give rise to additional contributions
that we need to subtract

\beq \phi^{(1)}_{ii} ( u, v ) = - \lambda \left( \frac{ 3 p^2 +
p^3 ( \Theta (v) - \Theta (u) )}{2\pi} \right)  \log \left(
\frac{2uv}{\theta} \right) \  \label{333} \eeq

We have arrived at the final formula:

\begin{eqnarray} \phi^{(1)}  & = &  \lambda \left( \frac{ 3 p^2 + p^3 ( \Theta (v) -
\Theta (u) )}{2\pi} \right) \left[ \theta \left( cos \left(
\frac{2uv}{\theta} \right) + {\cal C}i \left( \frac{2uv}{\theta}
\right) - \log \left( \frac{2uv}{\theta} \right) \right) + \right. \nonumber \\
& + &  \left. 2uv \left( {\cal S}i \left( \frac{2uv}{\theta}
\right) - \ {\cal S}i ( + \infty) \right) \right] \label{334}
\end{eqnarray}

Fortunately the divergent terms around the wave packets $uv \sim
0$ cancels out. Again this field has support primarily only in a
small region around the wave packets $ 0 < uv \lesssim \theta $,
apart from the oscillating cosine term and a logarithmic term
which is less divergent than the corresponding classical term for
$uv$ large, as in eq.  (\ref{38}). The asymptotic states are
modified permanently by noncommutativity also when the two wave
packets are far apart. It is probably this characteristic which
complicates the picture at the quantum level, since the $S$-matrix
approach is useful only in those cases where the interaction
switches off at large times.

\section{Conclusions}

In this preliminary investigation we have found an example of
scattering which hopefully can be solved at a nonperturbative
level. We have compared the classical scattering of kinks in
($1+1$) Higgs model with the noncommutative case. At a classical
level we have found a smooth solution without divergencies. This
solution is based on introducing an ansatz, which reduces the
equations of motions to a single nonlinear differential equation.
We have been able to have full control of it by combining analytic
and numerical methods. The solution of this equation is similar to
a Bessel function of order zero, which contains damped
oscillations towards a constant asymptotic value.

In the noncommutative case, the solution we have found is only
perturbative;  at this level there appear logarithmic terms which
are divergent both near the wavefronts and at infinity. However
the whole combination of terms cooperates to eliminate the
divergencies near the wavefronts. Instead the divergence at
infinity cannot be eliminated at a fixed order of perturbation
theory, but only resuming all orders of perturbation theory.

The peculiar characteristic of noncommutativity is to dress the
sharp wavefronts of the kinks giving them a size of order
$\sqrt{\theta}$ permanently, also when the wavefronts are far
apart. Noncommutativity modifies the asymptotic states in such a
way that the asymptotic states for two-body cannot be factorized
into a product of one-body states. These properties complicate the
quantization of such theories, since the $S$-matrix approach is
useful only for short range interactions switching off at large
times.

This work leaves open many questions; firstly it would be nice to
solve this model exactly at all orders and prove that there is a
smooth solution describing the scattering of noncommutative kinks.
Then investigating deeply the characteristics of the
noncommutative scattering we can look for the right axioms on
which to base the quantization of such theories. For example we
remember that there are many efforts to introduce an arrow of time
in quantum mechanics, by extending the ordinary Hilbert space into
a Rigged Hilbert space \cite{21}-\cite{24}. We personally believe
that only field theory, rather than one-particle quantum
mechanics, is able to produce microscopic irreversibility, because
it contains infinite degrees of freedom and thermodynamical
behaviours are possible only for systems with large number of
degrees of freedom.


\begin{thebibliography}{999}

\bibitem{1} B. Misra and I. Prigogine, " Irreversibility and
nonlocality ", Lett. Math. Phys. {\bf 7} (1983), 421.

\bibitem{2} T. Petrosky, I. Prigogine, " Poincare' resonances and
the extension of classical dynamics ", Chaos, Solitons and
Fractals {\bf 7} (1996), 441.

\bibitem{3} I. Prigogine, " Laws of Nature, probability and time
symmetry breaking ", Physica A {\bf 263} (1999), 528.

\bibitem{4} T. Petrosky, I. Prigogine, " Thermodynamical limit,
Hilbert space and breaking of time symmetry ", Chaos, Solitons and
Fractals, {\bf 11} (2000), 373.

\bibitem{5} S. Doplicher, K. Fredenhagen, J. E. Roberts, "
 The quantum structure of space-time at the Planck scale and quantum
 fields ", Commun.Math.Phys. {\bf 172} (1995), 187; hep-th/0303037.

\bibitem{6} S. Doplicher, " Space-time and fields: a quantum
texture ",  Karpacz 2001, New developments in fundamental
interaction theories 204; hep-th/0105251.

\bibitem{7} D. Bahns, S. Doplicher, K. Fredenhagen, G. Piacitelli,
" On the unitarity problem in space-time noncommutative theories
", Phys. Lett. {\bf B533} (2002),178; hep-th/0201222.

\bibitem{8} N. Seiberg and E. Witten, " String Theory and
Noncommutative geometry ", JHEP {\bf 9909} (1999) 032;
hep-th/9908142.

\bibitem{9} A. Connes, M. R. Douglas and A. Schwarz, "
Noncommutative geometry and Matrix theory: compactification on
Tori ", JHEP {\bf 9802} (1998) 003; hep-th/9711162.

\bibitem{10} P. Valtancoli, " Projectors for the fuzzy sphere "
Mod.Phys.Lett. {\bf A16} (2001) 639; hep-th/0101189.

\bibitem{11} P. Valtancoli, " Stability of the fuzzy sphere solution from matrix
model", Int. J. Mod. Phys. {\bf A18} (2003), 967; hep-th/0206075

\bibitem{12} P. Valtancoli, " Solitons for the fuzzy sphere from
matrix model",  Int. J. Mod. Phys. {\bf A18} (2003), 1107;
hep-th/0209117.

\bibitem{13} P. Valtancoli, " Noncommutative instantons on D = 2N planes from matrix
models", Int. J. Mod. Phys. {\bf A18} (2003) 1125; hep-th/0209118.

\bibitem{14} P. Valtancoli, " Projective modules over the fuzzy four sphere",
Mod. Phys. Lett. {\bf A17} (2002) 2189; hep-th/0210166.

\bibitem{15} P. Valtancoli, " Matrix model for noncommutative gravity and
gravitational instantons", to be published on Int. J. Mod. Phys.;
hep-th/0303119.

\bibitem{16} P. Valtancoli, " Gravity on a fuzzy sphere", to be published on
Int. J. Mod. Phys.; hep-th/0306065.

\bibitem{17} J. Gomis, T. Mehen, " Space-time noncommutative field
theories and unitarity ", Nucl. Phys. {\bf B591} (2000), 265;
hep-th/0005129.

\bibitem{18} J. Gomis, K. Kamimura, J. Llosa, " Hamiltonian
formalism for space-time noncommutative theories ", Phys. Rev.
{\bf D63} (2001), 045003; hep-th/0006235.

\bibitem{19} N. Seiberg, L. Susskind, N. Toumbas, " Space-time
noncommutativity and causality ", JHEP {\bf 0006} (2000), 044;
hep-th/0005015.

\bibitem{20} Luis Alvarez-Gaume, J.L.F. Barbon, R. Zwicky, " Remarks
on time space noncommutative field theories ", JHEP {\bf 0105}
(2001), 057; hep-th/0103069.

\bibitem{21} A. Bohm and H. Kaldass, " Rigged Hilbert space
resonances and time asymmetric quantum mechanics ";
quant-ph/9909081.

\bibitem{22} C. Schulte, R. Twarock, A. Bohm, " The Rigged Hilbert
space formulation of quantum mechanics and its implications for
irreversibility "; quant-ph/9505004.

\bibitem{23} R. C. Bishop, " The arrow of time in Rigged Hilbert
space quantum mechanics ";
http://philsci-archive.pitt.edu/archive/00000814/

\bibitem{24} I.E. Antoniou, M. Gadella, E. Karpov, I. Prigogine, G.
Pronko, " Gamov algebras ", Chaos, Solitons and Fractals {\bf 12}
(2001) 2757.



\end{thebibliography}
\end{document}